\documentclass{article}
\usepackage{dcase2023,amsmath,graphicx,url,times,booktabs,tabularx}
\usepackage{multirow}
\usepackage{float}

\title{Few-shot Bioacoustic event detection at the DCASE 2023 challenge}

\name{I. Nolasco$^{1}$,
      B. Ghani$^{2}$,
      S. Singh$^{1}$,
      E. Vida\~{n}a-Vila$^{3}$,
      H. Whitehead$^{10}$,
      E. Grout$^{4, 5}$,
      M.G. Emmerson$^{7}$,
      F. H. Jensen$^{8}$
      }
\secondlinename{	
        I. Kiskin$^{9}$,
       J. Morford$^{6}$,
       A. Strandburg-Peshkin$^{4, 5}$,
    L. Gill$^{11}$, 
      H. Pamu\l{}a$^{12}$, 
      V. Lostanlen$^{13}$, 
      D. Stowell$^{2, 14}$
      }
\address{
$^1$ Centre for Digital Music (C4DM), Queen Mary University of London, London, UK\\
$^{2}$ Naturalis Biodiversity Centre, Leiden, NL\\
$^3$ La Salle Campus Barcelona, Ramon Llull University, Barcelona, ES\\
$^4$ Dept. of Biology \& Centre for the Advanced Study of Collective Behaviour, University of Konstanz, DE\\
$^5$ Dept. for the Ecology of Animal Societies, Max Planck Institute of Animal Behavior, DE\\
$^{6}$ The Oxford Navigation group, Dept. of Zoology, Oxford University, Oxford, UK \\
$^{7}$ School of Biological and Behavioural Sciences, Queen Mary University of London, London, UK \\
$^{8}$ Biology Dept, Syracuse University, NY, USA\\
$^{9}$ Institute for People-Centred AI, FHMS,
University of Surrey, Surrey, UK\\
$^{10}$ School of Science, Engineering and Environment, University of Salford, Manchester, UK\\
$^{11}$ Landesbund f\"{u}r Vogel- und Naturschutz; Naturkundemuseum Bayern/BIOTOPIA Lab, DE\\
$^{12}$ AGH University of Science and Technology, Kraków, PL\\
$^{13}$ Nantes Université, École Centrale Nantes, CNRS, LS2N, UMR 6004, F-44000 Nantes, FR \\
$^{14}$ Tilburg University, Tilburg, NL
}
\begin{document}

\ninept
\maketitle

\begin{abstract}
Few-shot bioacoustic event detection consists in detecting sound events of specified types, in varying soundscapes, while having access to only a few examples of the class of interest. 
This task ran as part of the DCASE challenge for the third time this year with an evaluation set expanded to include new animal species, and a new rule: ensemble models were no longer allowed. 
The 2023 few-shot task received submissions from 6 different teams with F-scores reaching as high as 63\% on the evaluation set.  
Here we describe the task, focusing on describing the elements that differed from previous years. We also take a look back at past editions to describe how the task has evolved. Not only have the F-score results steadily improved (40\% to 60\% to 63\%), but the type of systems proposed have also become more complex. Sound event detection systems are no longer simple variations of the baselines provided: multiple few-shot learning methodologies are still strong contenders for the task.

\end{abstract}


\section{Introduction}
\label{sec:intro}
\begin{figure}[t]
    \centering
    \includegraphics[width=0.8\columnwidth]{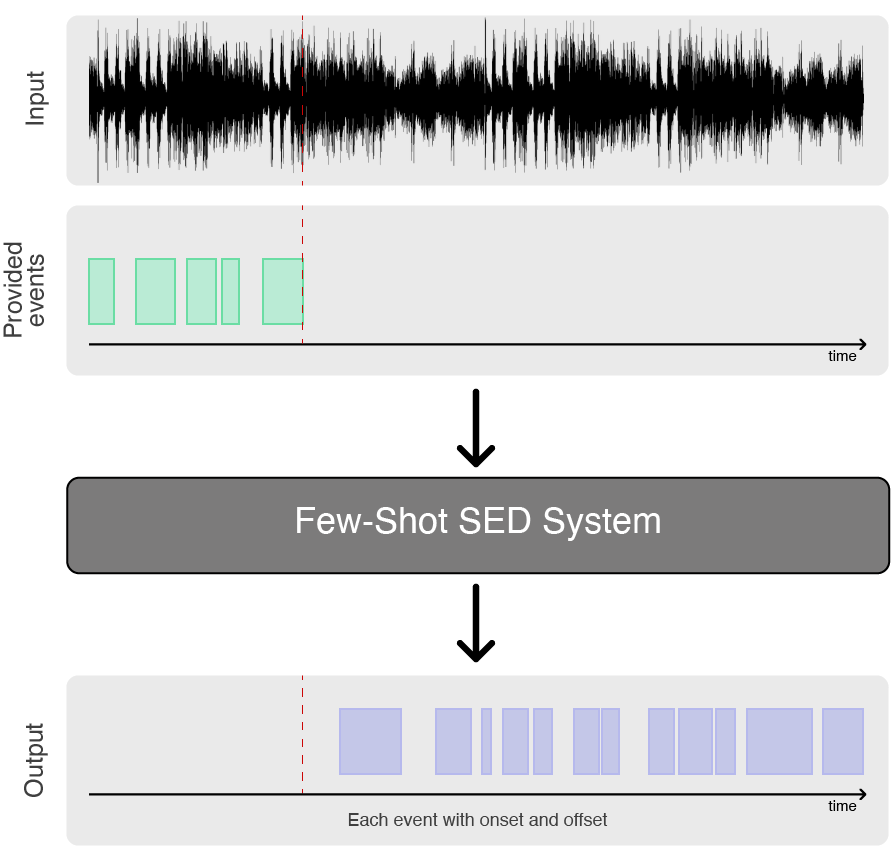}
    \caption{Overview of the proposed few-shot bioacoustic event detection task at the DCASE challenge. Green and purple rectangles represent labelled and predicted events, respectively.}
    \vspace{-0.5cm}
    \label{fig:overview}
\end{figure}

Bioacoustic event detection, the identification of animal vocalizations within specific timeframes, shares many similarities with sound event detection (SED) in varying contexts like urban settings~\cite{mesaros2010acoustic} or secured spaces~\cite{stowell2015detection, turpault2019sound}. Nonetheless, bioacoustics poses a unique set of challenges due to the varied recording conditions and diverse animal vocalizations~\cite{stowell2018computational}. This makes it an exciting and complex domain within machine learning, with several specialized sub-disciplines focused on different animals. Recent advances in supervised deep convolutional networks (CNNs) have potential for enhancing feature detection. However, their supervised nature necessitates extensive, well-categorized acoustic event data and hundreds of annotated examples per class. Gathering this data can be an uphill battle, considering the uneven distribution of species, the labor-intensive nature of audio annotation, and the variable taxonomy based on the use case~\cite{nolasco2023learning}.
\begin{table*}[h]
    \centering
    \begin{tabular}{r|l|c|c|c|c|c}
         & \textbf{Dataset} & \textbf{mic type} &  \textbf{\# audio files} & \textbf{total duration} & \textbf{\# labels} & \textbf{\# events} \\
         \hline
        \multirow{4}{*}{Training set} & BV: BirdVox-DCASE-10h & fixed & 5 & 10 hours & 11 & 9026\\
        & HT: Hyenas & various & 5& 5 hours & 5 & 611\\
        & MT: Meerkats & animal mounted & 2 & 70 mins & 4 & 1294\\
        & JD: Jackdaws & mobile & 1 & 10 mins & 1 & 357\\
        & WMW: Western Mediterranean Wetlands Birds & various & 161 & 5 hours  & 26 & 2941\\
        \hline
        \multirow{2}{*}{Validation set} & HB: Humbug mosquitoes& handheld & 10 & 2.38 hours & 1 & 712 \\
        & PB: Polish Baltic Sea bird flight calls & fixed &  6 & 3 hours & 2 & 292 \\
        & ME: Meerkats & animal mounted & 2 & 20 mins & 2 & 73\\
        \hline
        \hline
        \multirow{3}{*}{Evaluation Set} & CHE: Transfer-Exposure-Effects birds & fixed & 18 & 3 hours & 3 & 2550\\
        & DC: BIOTOPIA Dawn Chorus birds & fixed &  10 & 95 mins & 3 & 967\\
        & CT: Coati & handheld & 3  & 48 mins  & 3 & 365 \\
        & MS: Manx shearwater birds & fixed &  4 & 40 mins & 1 & 1087\\
        & QU: Dolphin quacks & animal mounted &  8 & 74 mins & 1 & 3441\\
        & MGE: Chick calls birds & fixed &  3 & 32 mins & 2 & 1195\\
        & CHE23: Transfer-Exposure-Effects Frogs & fixed & 16 & 40 mins & 1 & 798 \\
        & CW: Cow moos & fixed & 4 & 56 mins & 1 & 293 \\
        
    \end{tabular}
    \caption{Summary of dataset characteristics.}
    \label{tab:datasets}
\end{table*}
The limitations of a supervised sound event detection system become more prominent when extrapolating techniques used in speech to other animal sounds. This complexity arises from the differences in sound duration, interest units, and the context in which the sounds are made. Crucially, understanding the commencement and termination times of animal sounds is vital to community ecology, shedding light on various patterns of communication and influence among species~\cite{stowell2016detailed}. Unlike speech science with its relatively limited granularity, bioacoustic studies operate at multiple levels, from coarse classification of species to fine distinction of individual call types. Moreover, the diversity in recording equipment used for animal sounds, from far-field to underwater, adds another layer of complexity, transforming bioacoustic event detection into a collection of small-data problems, each requiring specialized systems. This fragmentation, although useful for species classification tasks, impedes the practical application of deep learning in bioacoustics and life sciences more broadly~\cite{nolasco2023learning}.
To address these challenges, this DCASE task proposes a unified approach for bioacoustic event detection across the various subdomains, aiming to mitigate the problems associated with data acquisition, annotation, and the fragmentation in computational bioacoustics. Hence, we compiled a unique ensemble of 14 small-scale datasets, each between 10 minutes and 10 hours long and derived from distinct sources, representing different application contexts. Breaking from the norm of training individual machine learning systems for each dataset, the idea is to develop a single, versatile system capable of identifying sound events across various datasets, with event categories specified at "query time". Additionally, during an evaluation on an audio file, the system is provided with the initial five instances of the desired sound event. This approach employs a machine learning paradigm known as "few-shot learning" (FSL)~\cite{snell2017prototypical, wang2021few}, where the aim is to construct precise models using less training data. In this context, FSL is explored using N-way-k-shot classification, where N and k represent the number of classes and the examples per class, respectively. Upon training with the first five occurrences of an event, the system effectively detects subsequent instances of the same event. Figure~\ref{fig:overview} provides an overview of the proposed task. Our hypothesis is that bioacoustic event detectors can be trained using available bioacoustic datasets and then generalized to new targets using a few examples at the time of deployment.

\section{Datasets}
\label{sec:format}
At the start of the DCASE challenge, each task releases its own Development set, consisting in a training and validation sets. Participants must use this dataset to develop and validate their systems. As the challenge enters the evaluation phase, the Evaluation set is released and participants apply their developed systems and output the predictions used to calculate the final ranking scores. 
These datasets are organised in subsets that represent different acoustic sources and were gathered here with the specific purpose of broadening the targeted species. A summary of the main characteristics are presented in Table \ref{tab:datasets}. Overall there are 8 sets focusing on bird species, 5 sets of mammal vocalisations (one of which underwater), 1 set of flying insect sounds (HB) and 1 set of amphibian calls (CHE23). 

For the Few shot bioacoustic task, the training set is multi-label, since the provided annotations contain more than one class of interest. However, both validation and evaluation sets are single label, meaning that each audio file is annotated only for a single class of interest. While events of other classes are present these are not annotated and should not be predicted by the systems.
Also, Given the few shot setup of this task, each audiofile of the evaluation set is accompanied only with the annotations for the 5 initial events of the class of interest.
The datasets used on the 2023 edition of the task remain the same as in previous edition, but the evaluation set has been extended with two new subsets of data: Cow moos (CW) and frog croakings (CHE23). 

\textbf{Cow moos (CW):} This dataset contains 4 audio files of about 15 minutes each recorded on a Cow's farm in Catalonia, Spain. An ambient microphone connected to a Zoom H5 recorder was hung on the ceiling of a yard with multiple cows. Cow vocalizations were recorded and manually labelled by researchers from La Salle Campus Barcelona and AWEC Advisors S.L. in the framework of the projects CowTalk and CowTalk-Pro. 

\textbf{Transfer-Exposure-Effects Frogs (CHE23): } This dataset is part of the same project which originated the CHE dataset, data were collected using unattended acoustic recorders (Songmeter 3) in the Chornobyl Exclusion Zone (CEZ) to capture the Chornobyl soundscape and investigate the longterm effects of the nuclear plower plant accident on the local ecology. 
The CHE23 dataset consists in 16 audiofiles of varying lengths annotated for frog croaking events, however many other calls of other species are present through out the recordings. 
The annotations were produced by Helen Whitehead using Raven Pro 1.6.

\begin{table*}[!h]
    \centering
    \begin{tabular}{l|c|l|l|l}
    \textbf{Team name} & 
    \multicolumn{1}{l|}{\begin{tabular}[c]{@{}l@{}}\textbf{Best} \\ \textbf{submission} \end{tabular}}   & \multicolumn{1}{l|}{\begin{tabular}[c]{@{}l@{}}\textbf{Eval set:} \\ \textbf{$F$-score \% (95\% CI)}\end{tabular}} 
    & \multicolumn{1}{l|}{\begin{tabular}[c]{@{}l@{}} \textbf{Val set} \\\textbf{ $F$-score \%} \end{tabular}} &
     \multicolumn{1}{l}{\begin{tabular}[c]{@{}l@{}} \textbf{Main characteristics} \end{tabular}} \\
    \hline
    Du\_NERCSLIP & 2 & 63.78 & 75.6  & Prototypical network with frame level embeddings; \\multitask learning; 
     & & & & Voice activity detection  \\
     \hline
    Moummad\_IMT & 2 & 42.72 & 63.46  & Contrastive learning learns an Embedding space;\\
    && & &fine-tuning encoder on both positive and negative events;\\
   \hline
    XuQianHu\_NUDT\_BIT &  3&42.5 &  63.94 & prototypical network, Delta MFCC and PCEN; \\
    & & & &Squeeze Excitation blocks \\
    \hline
    Gelderblom\_SINTEF &  2&31.10 &   & Encoder based on BEATs; prototypical network. \\
    \hline
    Jung\_KT & 3& 27.12 & 81.52  & Prototypical network trained with a Negative-based loss \\
    \hline
    Wilkinghoff\_FKIE &4 &  16.00 & 62.636  & Embeddings learnt with temporal dimension;\\
    & & & & template matching with Dynamic warping.  \\
    \hline
    
    \end{tabular}
    \caption{F-score results per team (best scoring system) on evaluation and validation sets, and summary of system characteristics. Systems are ordered by higher scoring rank on the evaluation set. These results and technical reports for the submitted systems can be found on task 5 results page \cite{task5resultspage}.}
    \label{tab:teams}
\end{table*} 

The remaining datasets re-used from the past editions have been thoroughly described in \cite{nolasco2023learning}.

\section{Baselines and Evaluation metrics }
\label{sec:pagelimit}

The benchmarks and evaluation metrics remain identical to those established in the 2022 rendition of the task~\cite{nolascofew}. The associated code can be procured from the GitHub repository\footnote{\url{https://github.com/c4dm/dcase-few-shot-bioacoustic}}.

The few-shot bioacoustic sound event detection task adopts two baselines: 1) Template matching, and 2) Protoypical networks. Template matching represents a common practice in the bioacoustics domain. The overall approach consists in taking each of the 5 initial examples as templates and cross-correlate each template with the remaining audiofile. Events are predicted by selecting time frames where the cross-correlation values surpasses a defined threshold. Prototypical networks~\cite{snell2017prototypical}, on the other hand are trained through episodic learning and employ a 5-way-2-shot classification model in our case. Each prototype represents a coordinate in vector space, calculated as a mean of the coordinates of the $5$ samples. Training comprises a \textit{Support set} of $5$ labelled samples from each class, while the remaining samples form the \textit{Query set}. A class prototype is computed via an embedding function with learning parameters. Distances are optimised, and the network training creates a general representation where similar sounds are closer. In this way, the future data points are labelled using nearest-neighbour algorithms.

The systems are evaluated based on how well they predict events on the evaluation set.  
The metric used combines intersection over union and bipartite graph matching algorithms to select the best matches between the predicted events and ground truth events. 
After the matching phase, we count the number of  true positives (TP) , false positives (FP), and false negatives (FN), which are then used to calculate precision, recall and F-score. The systems are ranked using the event based F-score value. The task description and details are provided in~\cite{nolasco2023learning}.

In this year's task, a distinctive modification has been introduced in the evaluation procedure. The use of ensemble models was no longer allowed. The objective behind this rule is to incentivise the development of truly general models, rather than a simple fusion of completely independent models. 

\section{Results}
\label{sec:pagestyle}

\begin{figure}[!b]
    \centering
    \includegraphics[width=\linewidth]{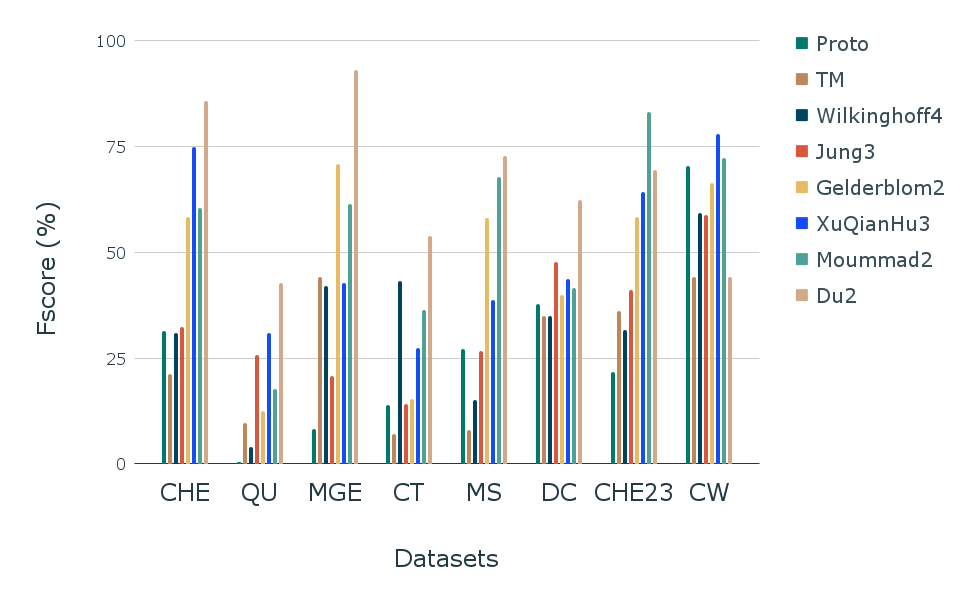}
    \caption{F-Score results by team (best submission only). Systems are ordered from least to  highest scoring rank on the evaluation set.}
    \label{fig:Fscore_dataset}
\end{figure}

The third edition of the Few-shot bioacoustic event detection task received participation of 6 teams, with a total of 22 submitted systems. 
The overall F-scores for the best submission per team are presented in Table~\ref{tab:teams} together with the main characteristics of the respective systems, and the results on each dataset of the evaluation set are presented in Fig.~\ref{fig:Fscore_dataset}.

\begin{figure*}[!h]
    \centering
    \includegraphics[width=0.8\linewidth]{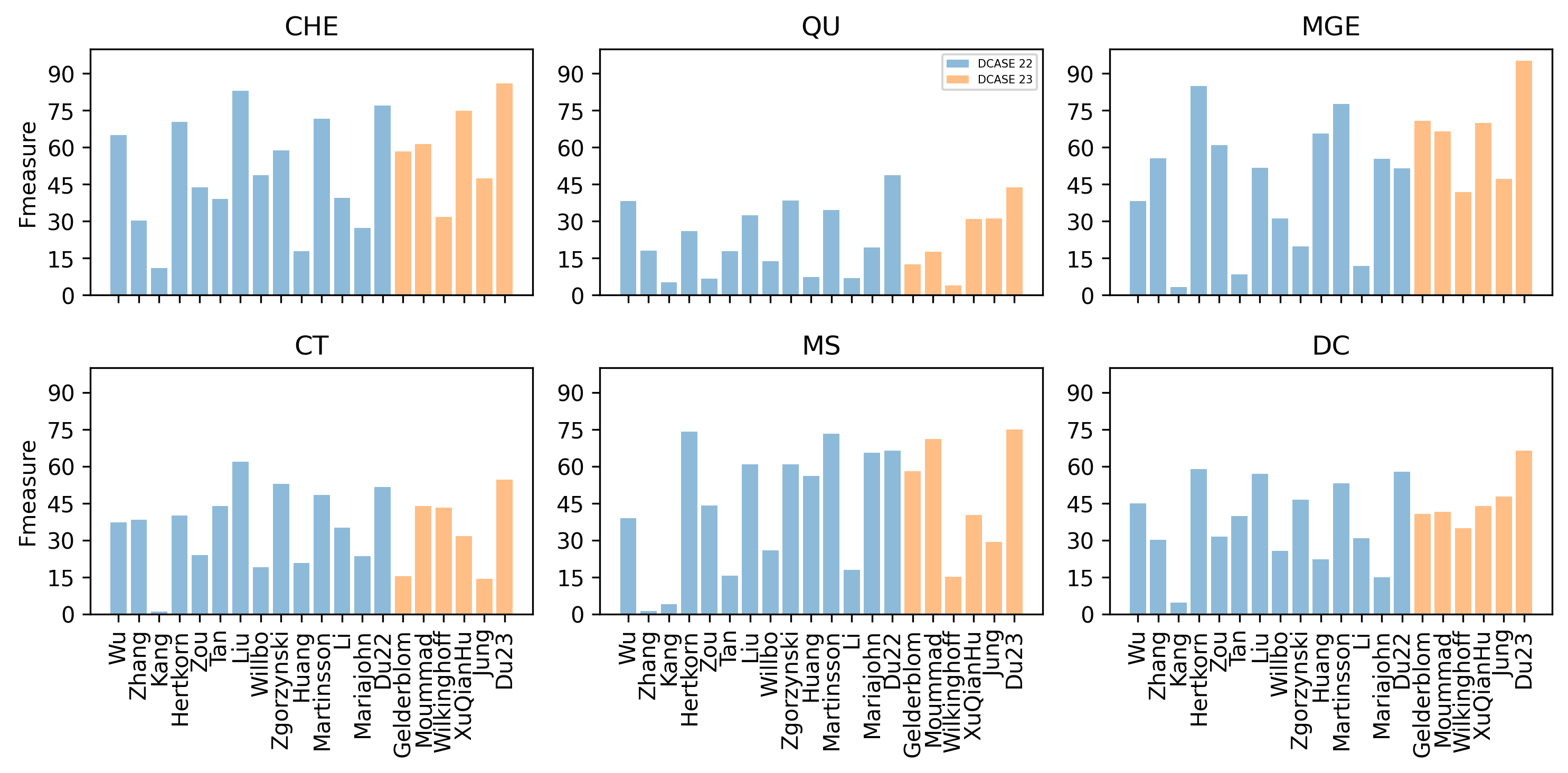}
    \caption[]{
    \setcounter{footnote}{1}
    Comparison of the maximum F-score achieved by each team for the datasets used in both 2022\footnotemark ~and 2023. The results for each year are distinguished by different colors.}
\label{fig:Fscore_compareprevious}
\end{figure*}

The winning submission is by the same team that won the previous edition of this task, namely, \textbf{Du\_NERCSLIP}. The system improves on the last year's submission, \cite{Du2022a} by including their frame-level embedding system into a multi-task learning architecture. The new system now includes Target Speaker Voice Activity Detection as one of the branches. This system achieved 63\% F-score which is an increase from the best F-score from last year, that was aprox. 60\%. Observing Fig.~\ref{fig:Fscore_compareprevious}, it is possible to observe the improved results on individual datasets for this team compared to the last year's system (columns Du22 and Du23).  This shows that the described modifications are responsible for the considerable increase in the overall F-score. 

Furthermore, an intriguing observation when looking at the F-scores per dataset in Fig.~\ref{fig:Fscore_dataset}, is that overall systems performed extremely well on the CW dataset, but not the winning submission. Indeed the performance of Du\_NERCSLIP's system on CW dataset is similar to the performance on the QU (doplphin quacks) dataset, which is considered to be an extremely difficult case due to its very short events. 

\footnotetext{\url{https://dcase.community/challenge2022/task-few-shot-bioacoustic-event-detection-results}}

\textbf{Moummad\_IMT} implemented a system based on Contrastive Learning, a method to learn an embedding space that maximises the distinction between positive events and negative events. 
During the evaluation stage, the encoder is further fine-tuned on the 5 POS examples provided and on selected NEG examples for each audiofile. The final predictions are then the result of a simple binary classification in this embedding space. 

\textbf{XuQianHu\_NUDT\_ BIT} largely improved upon the Prototypical Network baseline by including squeeze/excitation (SE) blocks into the encoder part of the network.  
The idea behind this is to create an adaptive mechanism that assigns different weights to different channels
of the feature map, by predicting their importance. 
The system also adopts the negative sample search mechanism proposed by Liu \textit{et al.} on the previous edition of the task~\cite{Liu2022a}, which is designed to improve the learning of the negative prototypes. 
Also following from Liu \textit{et al} submission, the input features employed are Delta MFCC and PCEN.

\textbf{Gelderblom\_SINTEF} followed the Prototypical Network approach, but use the BEATs pretrained model as encoder. 
BEAT stands for Bidirectional Encoder representation from Audio Transformers released by Microsoft for audio tokenisation and classification.
In their submission, the authors explore how useful this model is to represent bioacoustic data and compare the embeddings produced by the pretrained model with the embeddings produced after a few epochs of fine tuning on the ECS50 dataset.

\textbf{Jung\_KT} combines Contrastive Learning and Prototypical Networks.
This specifically addresses the problem that the high imbalance between positive samples and Negative samples creates in the learning of the prototypes. They propose a novel negative-based prototypical loss function that is used in a fine tuning stage of the pipeline and drives the system to maximise the positive to negative samples distance and minimise the distance between negative samples.

\textbf{Wilkinghoff\_FKIE} adopts template matching and dynamic time warping applied to embeddings trained with temporal resolution. The embedding model is trained to predict both class and temporal position of the sound event.

Observing the results spanning the two-year period (see Fig.~\ref{fig:Fscore_compareprevious}), it is evident that each dataset presents unique challenges for various algorithms. Notably, the QU dataset consistently proved to be difficult for all participating teams across both years. A comprehensive discussion on the last year's results, that could explain some of these results, is available in~\cite{nolasco2023learning}.

\section{Conclusion}
\label{sec:typestyle}
The 2023 edition of the Few-shot bioacoustic event detection task received some very innovative systems that reflect the state-of-the-art in Few-shot learning. 
We especially underscore the introduction of a novel technique, such as contrastive learning, making its initial entry in the history of the task's execution. Contrastive learning in the audio domain has seen increasing success and seems like a promising approach for the Few-shot problem.  

Also of note is the quality of the evaluation set gathered this year. The dataset now extends to 3 different taxonomic groups: mammals, birds and amphibians, which is a good indicator of the variety of challenges faced in the bioacoustics domain.

Moving forward we would be interested in analysing how exactly the characteristics of the different datasets impact each system and be able to understand if a single general model is indeed capable of predicting many different classes based on such few examples. The work in~\cite{nolasco2023learning} started to tackle these questions, and while it is still not clear, the improving results on successive editions of this task indicate that the Few-shot setting is a way to go. 

\bibliographystyle{IEEEtran}
\bibliography{refs}

\begin{thebibliography}{10}
\providecommand{\url}[1]{#1}
\def\UrlFont{\rmfamily}
\providecommand{\newblock}{\relax}
\providecommand{\bibinfo}[2]{#2}
\providecommand\BIBentrySTDinterwordspacing{\spaceskip=0pt\relax}
\providecommand\BIBentryALTinterwordstretchfactor{4}
\providecommand\BIBentryALTinterwordspacing{\spaceskip=\fontdimen2\font plus
\BIBentryALTinterwordstretchfactor\fontdimen3\font minus
  \fontdimen4\font\relax}
\providecommand\BIBforeignlanguage[2]{{%
\expandafter\ifx\csname l@#1\endcsname\relax
\typeout{** WARNING: IEEEtran.bst: No hyphenation pattern has been}%
\typeout{** loaded for the language `#1'. Using the pattern for}%
\typeout{** the default language instead.}%
\else
\language=\csname l@#1\endcsname
\fi
#2}}

\bibitem{mesaros2010acoustic}
A.~Mesaros, T.~Heittola, A.~Eronen, and T.~Virtanen, ``Acoustic event detection
  in real life recordings,'' in \emph{2010 18th European signal processing
  conference}.\hskip 1em plus 0.5em minus 0.4em\relax IEEE, 2010, pp.
  1267--1271.

\bibitem{stowell2015detection}
D.~Stowell, D.~Giannoulis, E.~Benetos, M.~Lagrange, and M.~D. Plumbley,
  ``Detection and classification of acoustic scenes and events,'' \emph{IEEE
  Transactions on Multimedia}, vol.~17, no.~10, pp. 1733--1746, 2015.

\bibitem{turpault2019sound}
N.~Turpault, R.~Serizel, A.~P. Shah, and J.~Salamon, ``Sound event detection in
  domestic environments with weakly labeled data and soundscape synthesis,'' in
  \emph{Workshop on Detection and Classification of Acoustic Scenes and
  Events}, 2019.

\bibitem{stowell2018computational}
D.~Stowell, ``Computational bioacoustic scene analysis,'' \emph{Computational
  analysis of sound scenes and events}, pp. 303--333, 2018.

\bibitem{nolasco2023learning}
I.~Nolasco, S.~Singh, V.~Morfi, V.~Lostanlen, A.~Strandburg-Peshkin,
  E.~Vida{\~n}a-Vila, L.~Gill, H.~Pamu{\l}a, H.~Whitehead, I.~Kiskin,
  \emph{et~al.}, ``Learning to detect an animal sound from five examples,''
  \emph{arXiv preprint arXiv:2305.13210}, 2023.

\bibitem{stowell2016detailed}
D.~Stowell, L.~Gill, and D.~Clayton, ``Detailed temporal structure of
  communication networks in groups of songbirds,'' \emph{Journal of the Royal
  Society Interface}, vol.~13, no. 119, p. 20160296, 2016.

\bibitem{snell2017prototypical}
J.~Snell, K.~Swersky, and R.~Zemel, ``Prototypical networks for few-shot
  learning,'' \emph{Advances in neural information processing systems},
  vol.~30, 2017.

\bibitem{wang2021few}
Y.~Wang, N.~J. Bryan, M.~Cartwright, J.~P. Bello, and J.~Salamon, ``Few-shot
  continual learning for audio classification,'' in \emph{ICASSP 2021-2021 IEEE
  International Conference on Acoustics, Speech and Signal Processing
  (ICASSP)}.\hskip 1em plus 0.5em minus 0.4em\relax IEEE, 2021, pp. 321--325.

\bibitem{task5resultspage}
\url{https://dcase.community/challenge2023/task-few-shot-bioacoustic-event-detection-results},
  accessed: 10-06-2023.

\bibitem{nolascofew}
I.~Nolasco, S.~Singh, E.~Vidana-Vila, E.~Grout, J.~Morford, M.~E.~F. Jensen,
  I.~Kiskin, H.~Whitehead, A.~Strandburg-Peshkin, L.~Gill10, \emph{et~al.},
  ``Few-shot bioacoustic event detection at the dcase 2022 challenge.''

\bibitem{Du2022a}
J.~Tang, Z.~Xueyang, T.~Gao, D.~Liu, X.~Fang, J.~Pan, Q.~Wang, J.~Du, K.~Xu,
  and Q.~Pan, ``Few-shot embedding learning and event filtering for bioacoustic
  event detection technical report,'' DCASE2022 Challenge, Tech. Rep., June
  2022.

\bibitem{Liu2022a}
H.~Liu, X.~Liu, X.~Mei, Q.~Kong, W.~Wang, and M.~D. Plumbley, ``Surrey system
  for dcase 2022 task 5 : Few-shot bioacoustic event detection with
  segment-level metric learning technical report,'' DCASE2022 Challenge, Tech.
  Rep., June 2022.

\end{thebibliography}

%
%
%
%
%
%
%
%
%

\end{document}